\documentclass[runningheads,orivec]{llncs}

\usepackage{xcolor}
\usepackage{textcomp}
\definecolor{P@Blue}{named}{blue}
\definecolor{P@ColorOnBlue}{gray}{.95}
\definecolor{P@GrayFG}{named}{darkgray}
\definecolor{P@GrayBG}{gray}{.90}
\definecolor{P@GrayComment}{gray}{.40}
\RequirePackage{listings}
\lstdefinelanguage{egison}{%
  sensitive = true,
  alsoletter={-},
  keywords = [1]{match-all, match-all-dfs, match, define, lambda},
  comment=[l]{;},
}%
\lstdefinelanguage{haskell}{%
  sensitive = true,
  comment=[l]{--},
}%
\usepackage{bussproofs}
\usepackage{amsmath}
\usepackage{amssymb}
\usepackage{enumitem}
\usepackage{url}
\usepackage{syntax}
\usepackage{graphicx}
\usepackage{authblk}
\usepackage{makecell,shortvrb}
\usepackage{multicol}
\usepackage{hhline}

\EnableBpAbbreviations

\newcommand*{\ppm}[3]{#1 \approx^{#2} #3 \Downarrow}
\newcommand*{\pdm}[2]{#1 \approx #2 \Downarrow}
\newcommand*{\mfun}[4]{#1 \sim^{#2}_{#3} #4 \Downarrow}
\newcommand*{\matom}[3]{#1 \sim_{#2} #3}
\newcommand*{\ev}[2]{#1, #2 \Downarrow}
\newcommand*{\set}[1]{\{#1\}}
\newcommand*{\cons}{:}
\newcommand*{\none}{\mathord{\texttt{none}}}
\newcommand*{\some}{\mathop{\texttt{some}}}
\newcommand*{\opt}{\mathop{\texttt{opt}}}

\let\emptyset\varnothing

\begin{document}

\title{Non-linear Pattern Matching with Backtracking for Non-free Data Types}
\author{Satoshi Egi\inst{1} \and Yuichi Nishiwaki\inst{2}}
\institute{Rakuten Institute of Technology, Japan \and University of Tokyo, Japan}

\maketitle

\begin{abstract}
\emph{Non-free data types} are data types whose data have no canonical forms.
For example, multisets are non-free data types because the multiset $\{a,b,b\}$ has two other equivalent but literally different forms $\{b,a,b\}$ and $\{b,b,a\}$.
\emph{Pattern matching} is known to provide a handy tool set to treat such data types.
Although many studies on pattern matching and implementations for practical programming languages have been proposed so far, we observe that none of these studies satisfy all the \emph{criteria of practical pattern matching}, which are as follows: i) efficiency of the backtracking algorithm for non-linear patterns, ii) extensibility of matching process, and iii) polymorphism in patterns.

This paper aims to design a new \emph{pattern-matching-oriented} programming language that satisfies all the above three criteria.
The proposed language features clean Scheme-like syntax and efficient and extensible pattern matching semantics.
This programming language is especially useful for the processing of complex non-free data types that not only include multisets and sets but also graphs and symbolic mathematical expressions.
We discuss the importance of our criteria of practical pattern matching and how our language design naturally arises from the criteria.
The proposed language has been already implemented and open-sourced as the Egison programming language.
\end{abstract}

\section{Introduction}\label{introduction}

Pattern matching is an important feature of programming languages featuring data abstraction mechanisms.
Data abstraction serves users with a simple method for handling data structures that contain plenty of complex information.
Using pattern matching, programs using data abstraction become concise, human-readable, and maintainable.
Most of the recent practical programming languages allow users to extend data abstraction e.g.\, by defining new types or classes, or by introducing new abstract interfaces.
Therefore, a good programming language with pattern matching should allow users to extend its pattern-matching facility akin to the extensibility of data abstraction.

Earlier, pattern-matching systems used to assume one-to-one correspondence between patterns and data constructors.
However, this assumption became problematic when one handles data types whose data have multiple representations.
To overcome this problem, Wadler proposed the pattern-matching system views~\cite{wadler1987views} that broke the symmetry between patterns and data constructors.
Views enabled users to pattern-match against data represented in many ways.
For example, a complex number may be represented either in polar or Cartesian form, and they are convertible to each other.
Using views, one can pattern-match a complex number internally represented in polar form with a pattern written in Cartesian form, and vice versa, provided that mutual transformation functions are properly defined.
Similarly, one can use the \texttt{Cons} pattern to perform pattern matching on lists with joins, where a list \texttt{[1,2]} can be either \texttt{(Cons 1 (Cons 2 Nil))} or \texttt{(Join (Cons 1 Nil) (Cons 2 Nil))}, if one defines a normalization function of lists with join into a sequence of \texttt{Cons}.

However, views require data types to have a distinguished canonical form among many possible forms.
In the case of lists with join, one can pattern-match with \texttt{Cons} because any list with join is canonically reducible to a list with join with the \texttt{Cons} constructor at the head.
On the other hand, for any list with join, there is no such canonical form that has \texttt{Join} at the head.
For example, the list \texttt{[1,2]} may be decomposed with \texttt{Join} into three pairs: \texttt{[]} and \texttt{[1,2]}, \texttt{[1]} and \texttt{[2]}, and \texttt{[1,2]} and \texttt{[]}.
For that reason, views do not support pattern matching of lists with join using the \texttt{Join} pattern.

Generally, data types without canonical forms are called \emph{non-free data types}.
Mathematically speaking, a non-free data type can be regarded as a quotient on a free data type over an equivalence.
An example of non-free data types is, of course, list with join: it may be viewed as a non-free data type composed of a (free) binary tree equipped with an equivalence between trees with the same leaf nodes enumerated from left to right, such as \texttt{(Join Nil (Cons 1 (Cons 2 Nil)))} $=$ \texttt{(Join (Cons 1 Nil) (Cons 2 Nil))}.
Other typical examples include sets and multisets, as they are (free) lists with obvious identifications.
Generally, as shown for lists with join, pattern matching on non-free data types yields multiple results.\footnote{In fact, this phenomenon that ``pattern matching against a single value yields multiple results'' does not occur for free data types. This is the unique characteristic of non-free data types.}
For example, multiset \texttt{\{1,2,3\}} has three decompositions by the \texttt{insert} pattern: \verb|insert(1,{2,3})|, \verb|insert(2,{1,3})|, and \verb|insert(3,{1,2})|.
Therefore, how to handle multiple pattern-matching results is an extremely important issue when we design a programming language that supports pattern matching for non-free data types.

On the other hand, \emph{pattern guard} is a commonly used technique for filtering such multiple results from pattern matching.
Basically, pattern guards are applied after enumerating all pattern-matching results.
Therefore, substantial unnecessary enumerations often occur before the application of pattern guards.
One simple solution is to break a large pattern into nested patterns to apply pattern guards as early as possible.
However, this solution complicates the program and makes it hard to maintain.
It is also possible to statically transform the program in the similar manner at the compile time.
However, it makes the compiler implementation very complex.
\emph{Non-linear pattern} is an alternative method for pattern guard.
Non-linear patterns are patterns that allow multiple occurrences of same variables in a pattern.
Compared to pattern guards, they are not only syntactically beautiful but also compiler-friendly.
Non-linear patterns are easier to analyze and hence can be implemented efficiently (Section~\ref{motivation-non-linear} and~\ref{egison-non-linear}).
However, it is not obvious how to extend a non-linear pattern-matching system to allow users to define an algorithm to decompose non-free data types.
In this paper, we introduce \emph{extensible pattern matching} to remedy this issue (Section~\ref{motivation-user-defined},~\ref{egison-extensible}, and~\ref{matcher}).
Extensibility of pattern matching also enables us to define \emph{predicate patterns}, which are typically implemented as a built-in feature (e.g.\,pattern guards) in most pattern-matching systems.
Additionally, we improve the usability of pattern matching for non-free data types by introducing a syntactic generalization for the \texttt{match} expression, called \emph{polymorphic patterns} (Section~\ref{motivation-polymorphism} and~\ref{egison-polymorphism}).
We also present a non-linear pattern-matching algorithm specialized for backtracking on infinite search trees and supports pattern matching with infinitely many results in addition to keeping efficiency (Section~\ref{algorithm}).

This paper aims to design a programming language that is oriented toward pattern matching for non-free data types.
We summarize the above argument in the form of three criteria that must be fulfilled by a language in order to be used in practice:
\begin{enumerate}
\item Efficiency of the backtracking algorithm for non-linear patterns,
\item Extensibility of pattern matching, and
\item Polymorphism in patterns.
\end{enumerate}
We believe that the above requirements, called together \emph{criteria of practical pattern matching}, are fundamental for languages with pattern matching.
However, none of the existing languages and studies \cite{viewsWeb,erwig1996active,tullsen2000first,antoy2010functional} fulfill all of them.
In the rest of the paper, we present a language which satisfies the criteria, together with comparisons with other languages, several working examples, and formal semantics.
We emphasize that our proposal has been already implemented in Haskell as the \emph{Egison} programming language, and is open-sourced~\cite{egison}.
Since we set our focus in this paper on the design of the programming language, detailed discussion on the implementation of Egison is left for future work.

\section{Related Work}\label{related}

In this section, we compare our study with the prior work.

First, we review previous studies on pattern matching in functional programming languages.
Our proposal can be considered as an extension of these studies.

The first non-linear pattern-matching system was the symbol manipulation system proposed by McBride~\cite{mcbride1969symbol}.
This system was developed for Lisp.
Their paper demonstrates some examples that process symbolic mathematical expressions to show the expressive power of non-linear patterns.
However, this approach does not support pattern matching with multiple results, and only supports pattern matching against a list as a collection.

Miranda laws~\cite{turner1985miranda,thompson1986laws,thompson1990lawful} and Wadler's views~\cite{wadler1987views,okasaki1998views} are seminal work.
These proposals provide methods to decompose data with multiple representations by explicitly declaring transformations between each representation.
These are the earliest studies that allow users to customize the execution process of pattern matching.
However, the pattern-matching systems in these proposals treat neither multiple pattern matching results nor non-linear patterns.
Also, these studies demand a canonical form for each representation.

Active patterns~\cite{erwig1996active,syme2007extensible} provides a method to decompose non-free data.
In active patterns, users define a \textit{match function} for each pattern to specify how to decompose non-free data.
For example, \texttt{insert} for multisets is defined as a match function in \cite{erwig1996active}.
An example of pattern matching against graphs using matching function is also shown in \cite{erwig1997functional}.
One limitation of active patterns is that it does not support backtracking in the pattern matching process.
In active patterns, the values bound to pattern variables are fixed in order from the left to right of a pattern.
Therefore, we cannot write non-linear patterns that requires backtracking such as a pattern that matches with a collection (like sets or multisets) that contains two identical elements.
(The pattern matching fails if we unfortunately pick an element that appears more than twice at the first choice.)

First-class patterns~\cite{tullsen2000first} is a sophisticated system that treats patterns as first-class objects.
The essence of this study is a \textit{pattern function} that defines how to decompose data with each data constructor.
First-class patterns can deal with pattern matching that generates multiple results.
To generate multiple results, a pattern function returns a list.
A critical limitation of this proposal is that first-class patterns do not support non-linear pattern matching.

Next, we explain the relation with logic programming.

We have mentioned that non-linear patterns and backtracking are important features to extend the efficiency and expressive power of pattern matching especially on non-free data types.
Unification of logic programming has both features.
However, how to integrate non-determinism of logic programming and pattern matching is not obvious~\cite{hanus2007multi}.
For example, the pattern-matching facility of Prolog is specialized only for algebraic data types.

Functional logic programming~\cite{antoy2010functional} is an approach towards this integration.
It allows both of non-linear patterns and multiple pattern-matching results.
The key difference between the functional logic programming and our approach is in the method for defining pattern-matching algorithms.
In functional logic programming, we describe the pattern-matching algorithm for each pattern in the logic-programming style.
A function that describes such an algorithm is called a \emph{pattern constructor}.
A pattern constructor takes decomposed values as its arguments and returns the target data.
On the other hand, in our proposal, pattern constructors are defined in the functional-programming style: pattern constructors take a target datum as an argument and returns the decomposed values.
This enables direct description of algorithms.

\section{Motivation}\label{motivation}

In this section, we discuss the requirements for programming languages to establish practical pattern matching for non-free data types.

\subsection{Pattern Guards vs. Non-linear Patterns}\label{motivation-non-linear}

Compared to pattern guards, non-linear patterns are a compiler-friendly method for filtering multiple matching results efficiently.
However, non-linear pattern matching is typically implemented by converting them to pattern guards.
For example, some implementations of functional logic programming languages convert non-linear patterns to pattern guards \cite{antoy2010programming,antoy2001constructor,hanus2007multi}.
This method is inefficient because it leads to enumerating unnecessary candidates.
In the following program in Curry, \texttt{seqN} returns \verb|"Matched"| if the argument list has a sequential \texttt{N}-tuple.
Otherwise it returns \verb|"Not matched"|.
\texttt{insert} is used as a pattern constructor for decomposing data into an element and the rest ignoring the order of elements.
\begin{lstlisting}[language=haskell]
insert x [] = [x]
insert x (y:ys) = x:y:ys ? y:(insert x ys)

seq2 (insert x (insert (x+1) _)) = "Matched"
seq2 _ = "Not matched"

seq3 (insert x (insert (x+1) (insert (x+2) _))) = "Matched"
seq3 _ = "Not matched"

seq4 (insert x (insert (x+1) (insert (x+2) (insert (x+3) _)))) = "Matched"
seq4 _ = "Not matched"

seq2 (take 10 (repeat 0)) -- returns "Not matched" in O(n^2) time
seq3 (take 10 (repeat 0)) -- returns "Not matched" in O(n^3) time
seq4 (take 10 (repeat 0)) -- returns "Not matched" in O(n^4) time
\end{lstlisting}
When we use a Curry compiler such as PAKCS~\cite{pakcs} and KiCS2~\cite{brassel2011kics2}, we see that ``\verb|seq4 (take n (repeat 0))|'' takes more time than ``\verb|seq3 (take n (repeat 0))|'' because \verb|seq3| is compiled to \verb|seq3'| as follows.
Therefore, \verb|seq4| enumerates $\binom{n}{4}$ candidates, whereas \verb|seq3| enumerates $\binom{n}{3}$ candidates before filtering the results.
If the program uses non-linear patterns as in \verb|seq3|, we easily find that we can check no sequential triples or quadruples exist simply by checking $\binom{n}{2}$ pairs.
However, such information is discarded during the program transformation into pattern guards.
\begin{lstlisting}
seq3' (insert x (insert y (insert z _))) | y==x+1 && z==x+2 = "Matched"
seq3' _ = "Not matched"
\end{lstlisting}
One way to make this program efficient in Curry is to stop using non-linear patterns and instead use a predicate explicitly in pattern guards.
The following illustrates such a program.
\begin{lstlisting}
isSeq2 (x:y:rs) = y == x+1
isSeq3 (x:rs) = isSeq2 (x:rs) && isSeq2 rs

perm [] = []
perm (x:xs) = insert x (perm xs)

seq3 xs | isSeq3 ys = "Matched" where ys = perm xs
seq3 _ = "Not matched"

seq3 (take 10 (repeat 0))  -- returns "Not matched" in O(n^2) time
\end{lstlisting}
In the program, because of the laziness, only the head part of the list is evaluated.
In addition, because of \emph{sharing}~\cite{fischer2009purely}, the common head part of the list is pattern-matched only once.
Using this call-by-need-like strategy enables efficient pattern matching on sequential n-tuples.
However, this strategy sacrifices readability of programs and makes the program obviously redundant.
In this paper, instead, we base our work on non-linear patterns and attempt to improve its usability keeping it compiler-friendly and syntactically clean.

\subsection{Extensible Pattern Matching}\label{motivation-user-defined}

As a program gets more complicated, data structures involved in the program get complicated as well.
A pattern-matching facility for such data structures (e.g.\, graphs and mathematical expressions) should be extensible and customizable by users because it is impractical to provide the data structures for these data types as built-in data types in general-purpose languages. 


In the studies of computer algebra systems, efficient non-linear pattern-matching algorithms for mathematical expressions that avoid such unnecessary search have already been proposed~\cite{wolframPatternTutorial,krebber2017non}.
Generally, users of such computer algebra systems control the pattern-matching method for mathematical expressions by specifying attributes for each operator.
For example, the \texttt{Orderless} attribute of the Wolfram language indicates that the order of the arguments of the operator is ignored~\cite{wolframOrderless}.
However, the set of attributes available is fixed and cannot be changed~\cite{wolframAttributes}.
This means that the pattern-matching algorithms in such computer algebra systems are specialized only for some specific data types such as multisets.
However, there are a number of data types we want to pattern-match other than mathematical expressions, like unordered pairs, trees, and graphs.

Thus, extensible pattern matching for non-free data types is necessary for handling complicated data types such as mathematical expressions.
This paper designs a language that allows users to implement efficient backtracking algorithms for general non-free data types by themselves.
It provides users with the equivalent power to adding new attributes freely by themselves.
We discuss this topic again in Section~\ref{egison-extensible}.

\subsection{Monomorphic Patterns vs Polymorphic Patterns}\label{motivation-polymorphism}

Polymorphism of patterns is useful for reducing the number of names used as pattern constructors.
If patterns are monomorphic, we need to use different names for pattern constructors with similar meanings.
As such, monomorphic patterns are error-prone.

For example, the pattern constructor that decomposes a collection into an element and the rest ignoring the order of the elements is bound to the name \texttt{insert} in the sample code of Curry \cite{antoy2010programming} as in Section~\ref{motivation-non-linear}.
The same pattern constructor's name is \texttt{Add'} in the sample program of Active Patterns \cite{erwig1996active}.
However, these can be considered as a generalized \texttt{cons} pattern constructor for lists to multisets, because they are same at the point that both of them are a pattern constructor that decomposes a collection into an element and the rest.

Polymorphism is important, especially for value patterns.
A value pattern is a pattern that matches when the value in the pattern is equal to the target.
It is an important pattern construct for expressing non-linear patterns.
If patterns are monomorphic, we need to prepare different notations for value patterns of different data types.
For example, we need to have different notations for value patterns for lists and multisets.
This is because equivalence of objects as lists and multisets are not equal although both lists and multisets are represented as a list.
\begin{lstlisting}[language=egison]
pairsAsLists (insert x (insert x _)) = "Matched"
pairsAsLists _ = "Not matched"

pairsAsMultisets (insert x (insert y _)) | (multisetEq x y) = "Matched"
pairsAsMultisets _ = "Not matched"

pairsAsLists [[1,2],[2,1]]     -- returns "Not matched"
pairsAsMultisets [[1,2],[2,1]] -- returns "Matched"
\end{lstlisting}

\section{Proposal}

In this section, we introduce our pattern-matching system, which satisfies all requirements shown in Section~\ref{motivation}.
Our language has Scheme-like syntax.
It is dynamically typed, and as well as Curry, based on lazy evaluation.

\subsection{The \texttt{match-all} and \texttt{match} expressions}\label{egison-expr}

We explain the \texttt{match-all} expression.
It is a primitive syntax of our language.
It supports pattern matching with multiple results.

We show a sample program using \texttt{match-all} in the following.
In this paper, we show the evaluation result of a program in the comment that follows the program.
``\texttt{;}'' is the inline comment delimiter of the proposed language.
\begin{lstlisting}[language=egison]
(match-all {1 2 3} (list integer) [<join $xs $ys> [xs ys]])
; {[{} {1 2 3}] [{1} {2 3}] [{1 2} {3}] [{1 2 3} {}]}
\end{lstlisting}
Our language uses three kinds of parenthesis in addition to ``\texttt{(}'' and ``\texttt{)}'', which denote function applications.
``\texttt{<}'' and ``\texttt{>}'' are used to apply pattern and data constructors.
In our language, the name of a data constructor starts with uppercase, whereas the name of a pattern constructor starts with lowercase.
``\texttt{[}'' and ``\texttt{]}'' are used to build a tuple.
``\verb|{|'' and ``\verb|}|'' are used to denote a \emph{collection}.

In our implementation, the collection type is a built-in data type implemented as a lazy 2-3 finger tree~\cite{hinze2006finger}.
This reason is that we thought data structures that support a wider range of operations for decomposition are more suitable for our pattern-matching system. (2-3 finger trees support efficient extraction of the last element.)

\texttt{match-all} is composed of an expression called \textit{target}, \textit{matcher}, and \textit{match clause}, which consists of a \textit{pattern} and \textit{body expression}.
The \texttt{match-all} expression evaluates the body of the match clause for each pattern-matching result and returns a (lazy) collection that contains all results.
In the above code, we pattern-match the target \verb|{1 2 3}| as a list of integers using the pattern \texttt{<join \$xs \$ys>}.
\texttt{(list integer)} is a matcher to pattern-match the pattern and target as a list of integer.
The pattern is constructed using the \texttt{join} pattern constructor.
\texttt{\$xs} and \texttt{\$ys} are called \textit{pattern variables}.
We can use the result of pattern matching referring to them.
A \texttt{match-all} expression first consults the matcher on how to pattern-match the given target and the given pattern.
Matchers know how to decompose the target following the given pattern and enumerate the results, and \texttt{match-all} then collects the results returned by the matcher.
In the sample program, given a \texttt{join} pattern, \texttt{(list integer)} tries to divide a collection into two collections.
The collection \verb|{1 2 3}| is thus divided into two collections by four ways.

\texttt{match-all} can handle pattern matching that may yield infinitely many results.
For example, the following program extracts all twin primes from the infinite list of prime numbers\footnote{We will explain the meaning of the value pattern \texttt{,(+ p 2)} and the \texttt{cons} pattern constructor in Section~\ref{egison-non-linear} and~\ref{egison-polymorphism}, respectively.}.
We will discuss this mechanism in Section~\ref{mechanismInf}.
\begin{lstlisting}[language=egison]
(define $twin-primes
  (match-all primes (list integer)
    [<join _ <cons $p <cons ,(+ p 2) _>>> [p (+ p 2)]]))

(take 6 twin-primes) ; {[3 5] [5 7] [11 13] [17 19] [29 31] [41 43]}
\end{lstlisting}

There is another primitive syntax called \texttt{match} expression.
While \texttt{match-all} returns a collection of all matched results, \texttt{match} short-circuits the pattern matching process and immediately returns if any result is found.
Another difference from \texttt{match-all} is that it can take multiple match clauses.
It tries pattern matching starting from the head of the match clauses, and tries the next clause if it fails.
Therefore, \texttt{match} is useful when we write conditional branching.

However, \texttt{match} is inessential for our language.
It is implementable in terms of the \texttt{match-all} expression and macros.
The reason is because the \texttt{match-all} expression is evaluated lazily, and, therefore, we can extract the first pattern-matching result from \texttt{match-all} without calculating other pattern-matching results simply by using \texttt{car}.
We can implement \texttt{match} by combining the \texttt{match-all} and \texttt{if} expressions using macros.
Furthermore, \texttt{if} is also implementable in terms of the \texttt{match-all} and \texttt{matcher} expression as follows.
We will explain the \texttt{matcher} expression in Section~\ref{matcher}.
For that reason, we only discuss the \texttt{match-all} expression in the rest of the paper.

\begin{lstlisting}[language=egison]
(define $if
  (macro [$b $e1 $e2]
    (car (match-all b (matcher {[$ something {[<True>  {e1}]
                                              [<False> {e2}]}]})
           [$x x]))))
\end{lstlisting}

\subsection{Efficient Non-linear Pattern Matching with Backtracking}\label{egison-non-linear}

Our language can handle non-linear patterns efficiently.
For example, the calculation time of the following code does not depend on the pattern length.
Both of the following examples take $O(n^2)$ time to return the result.
\begin{lstlisting}[language=egison]
(match-all (take n (repeat 0)) (multiset integer)
  [<insert $x <insert ,(+ x 1) _>> x])
; returns {} in O(n^2) time

(match-all (take n (repeat 0)) (multiset integer)
  [<insert $x <insert ,(+ x 1) <insert ,(+ x 2) _>>> x])
; returns {} in O(n^2) time
\end{lstlisting}
In our proposal, a pattern is examined from left to right in order, and the binding to a pattern variable can be referred to in its right side of the pattern.
In the above examples, the pattern variable \texttt{\$x} is bound to any element of the collection since the pattern constructor is \texttt{insert}.
After that, the patterns ``\texttt{,(+ x 1)}'' and ``\texttt{,(+ x 2)}'' are examined.
A pattern that begins with ``\texttt{,}'' is called a \textit{value pattern}.
The expression following ``\texttt{,}'' can be any kind of expressions.
The value patterns match with the target data if the target is equal to the content of the pattern.
Therefore, after successful pattern matching, \texttt{\$x} is bound to an element that appears multiple times.

We can more elaborately discuss the difference of efficiency of non-linear patterns and pattern guards in general cases.
The time complexity involved in pattern guards is $O(n^{p+v})$ when the pattern matching fails, whereas the time complexity involved in non-linear patterns is $O(n^{p+min(1,v)})$, where $n$ is the size of the target object\footnote{Here, we suppose that the number of decompositions by each pattern constructor can be approximated by the size of the target object.}, $p$ is the number of pattern variables, and $v$ is the number of value patterns.
The difference between $v$ and $min(1,v)$ comes from the mechanism of non-linear pattern matching that backtracks at the first mismatch of the value pattern.


\begin{table}[htbp]
\begin{tabular}{llll}
\begin{tabular}{|l||l|l|l|l|l|} \hline
Curry & n=15 & n=25 & n=30 & n=50 & n=100 \\ \hhline{|=#=|=|=|=|=|}
\texttt{seq2} & 1.18s & 1.20s & 1.29s & 1.53s & 2.54s \\ \hline
\texttt{seq3} & 1.42s & 2.10s & 2.54s & 7.40s & 50.66s \\ \hline
\texttt{seq4} & 3.37s & 16.42s & 34.19s & 229.51s & 3667.49s \\ \hline
\end{tabular}
& & &
\begin{tabular}{|l||l|l|l|l|l|} \hline
Egison & n=15 & n=25 & n=30 & n=50 & n=100 \\ \hhline{|=#=|=|=|=|=|}
\texttt{seq2} & 0.26s & 0.34s &  0.43s & 0.84s &  2.72s \\ \hline
\texttt{seq3} & 0.25s & 0.34s &  0.46s & 0.82s &  2.66s \\ \hline
\texttt{seq4} & 0.25s & 0.34s & 0.42s & 0.78s & 2.47s \\ \hline
\end{tabular}
\end{tabular}
\caption{Benchmarks of Curry (PAKCS version 2.0.1 and Curry2Prolog(swi 7.6) compiler environment) and Egison (version 3.7.12)}
\label{table:benchmark}
\end{table}

Table~\ref{table:benchmark} shows micro benchmark results of non-linear pattern matching for Curry and Egison.
The table shows execution times of the Curry program presented in Section~\ref{motivation-non-linear} and the corresponding Egison program as shown above.
The environment we used was Ubuntu on VirtualBox with 2 processors and 8GB memory hosted on MacBook Pro (2017) with 2.3 GHz Intel Core i5 processor.
We can see that the execution times in two implementations follow the theoretical computational complexities discussed above.
We emphasize that this benchmark results do not mean Curry is slower than Egison.
We can write the efficient programs for the same purpose in Curry if we do not persist in using non-linear patterns.
Let us also note that the current implementation of Egison is not tuned up and comparing constant times in two implementations is nonsense.

Value patterns are not only efficient but also easy to read once we are used to them because it enables us to read patterns in the same order the execution process of pattern matching goes.
It also reduces the number of new variables introduced in a pattern.
We explain the mechanism how the proposed system executes the above pattern matching efficiently in Section~\ref{algorithm}.

\subsection{Polymorphic Patterns}\label{egison-polymorphism}

The characteristic of the proposed pattern-matching expression is that they take a matcher.
This ingredient allows us to use the same pattern constructors for different data types.

For example, one may want to pattern-match a collection \verb|{1 2 3}| sometimes as a list and other times as a multiset or a set.
For these three types, we can naturally define similar pattern-matching operations.
One example is the \texttt{cons} pattern, which is also called \texttt{insert} in Section~\ref{motivation-non-linear} and \ref{egison-non-linear}.
Given a collection, pattern \texttt{<cons \$x \$rs>} divides it into the ``head'' element and the rest.
When we use the \texttt{cons} pattern for lists, it either yields the result which is uniquely determined by the constructor, or just fails when the list is empty.
On the other hand, for multisets, it non-deterministically chooses an element from the given collection and yields many results.
By explicitly specifying which matcher is used in match expressions, we can uniformly write such programs in our language:
\begin{lstlisting}[language=egison]
(match-all {1 2 3} (list integer) [<cons $x $rs> [x rs]])
; {[1 {2 3}]}
(match-all {1 2 3} (multiset integer) [<cons $x $rs> [x rs]])
; {[1 {2 3}] [2 {1 3}] [3 {1 2}]}
(match-all {1 2 3} (set integer) [<cons $x $rs> [x rs]])
; {[1 {1 2 3}] [2 {1 2 3}] [3 {1 2 3}]}
\end{lstlisting}
In the case of lists, the head element \texttt{\$x} is simply bound to the first element of the collection.
On the other hand, in the case of multisets or sets, the head element can be any element of the collection because we ignore the order of elements.
In the case of lists or multisets, the rest elements \texttt{\$rs} are the collection that is made by removing the ``head'' element from the original collection.
However, in the case of sets, the rest elements are the same as the original collection because we ignore the redundant elements.
If we interpret a set as a collection that contains infinitely many copies of an each element, this specification of \texttt{cons} for sets is natural.
This specification is useful, for example, when we pattern-match a graph as a set of edges and enumerate all paths with some fixed length including cycles without redundancy.

Polymorphic patterns are useful especially when we use value patterns.
As well as other patterns, the behavior of value patterns is dependent on matchers.
For example, an equality \texttt{\{1 2 3\} = \{2 1 3\}} between collections is false if we regard them as mere lists but true if we regard them as multisets.
Still, thanks to polymorphism of patterns, we can use the same syntax for both of them.
This greatly improves the readability of the program and makes programming with non-free data types easy.
\begin{lstlisting}[language=egison]
(match-all {1 2 3} (list integer) [,{2 1 3} "Matched"]) ; {}
(match-all {1 2 3} (multiset integer) [,{2 1 3} "Matched"]) ; {"Matched"}
\end{lstlisting}

We can pass matchers to a function because matchers are first-class objects.
It enables us to utilize polymorphic patterns for defining function.
The following is an example utilizing polymorphism of value patterns.
\begin{lstlisting}[language=egison]
(define $member?/m
  (lambda [$m $x $xs]
    (match xs (list m) {[<join _ <cons ,x _>> #t] [_ #f]})))
\end{lstlisting}

\subsection{Extensible Pattern Matching}\label{egison-extensible}

In the proposed language, users can describe methods for interpreting patterns in the definition of matchers.
Matchers appeared up to here are defined in our language.
We show an example of a matcher definition.
We will explain the details of this definition in Section~\ref{unordered-pair}.
\begin{lstlisting}[language=egison]
(define $unordered-pair
  (lambda [$a]
    (matcher {[<pair $ $> [a a] {[<Pair $x $y> {[x y] [y x]}]}]
              [$ [something] {[$tgt {tgt}]}]})))
\end{lstlisting}
An \emph{unordered pair} is a pair ignoring the order of the elements.
For example, \texttt{<Pair 2 5>} is equivalent to \texttt{<Pair 5 2>}, if we regard them as unordered pairs.
Therefore, datum \texttt{<Pair 2 5>} is successfully pattern-matched with pattern \texttt{<pair ,5 \$x>}.
\begin{lstlisting}[language=egison]
(match-all <Pair 2 5> (unordered-pair integer) [<pair ,5 $x> x]) ; {2}
\end{lstlisting}
We can define matchers for more complicated data types.
For example, Egi constructed a matcher for mathematical expressions for building a computer algebra system on our language~\cite{egisonMath,egi2017scalar,egi2018scalar}.
His computer algebra system is implemented as an application of the proposed pattern-matching system.
The matcher for mathematical expressions is used for implementing simplification algorithms of mathematical expressions.
A program that converts a mathematical expression object $n\cos^2(\theta) + n\sin^2(\theta)$ to $n$ can be implemented as follows.
(Here, we introduced the \texttt{math-expr} matcher and some syntactic sugar for patterns.)
\begin{lstlisting}[language=egison]
(define $rewrite-rule-for-cos-and-sin-poly
  (lambda [$poly]
    (match poly math-expr
      {[<+ <* $n <,cos $x>^,2 $y> <* ,n <,sin ,x>^,2 ,y> $r>
        (rewrite-rule-for-cos-and-sin-poly <+' r <*' n y>>)]
       [_ poly]})))
\end{lstlisting}

\section{Algorithm}\label{algorithm}

This section explains the pattern-matching algorithm of the proposed system.
The formal definition of the algorithm is given in Section~\ref{formal-semantics}.
The method for defining matchers explained in Section~\ref{matcher} is deeply related to the algorithm.

\subsection{Execution Process of Non-linear Pattern Matching}

Let us show what happens when the system evaluates the following pattern-matching expression.
\begin{lstlisting}[language=egison]
(match-all {2 8 2} (multiset integer) [<cons $m <cons ,m _>> m]) ; {2 2}
\end{lstlisting}
Figure~\ref{fig:reductionPath} shows one of the execution paths that reaches a matching result.
First, the initial \emph{matching state} is generated (step 1).
A matching state is a datum that represents an intermediate state of pattern matching.
A matching state is a compound type consisting of a stack of \emph{matching atoms}, an environment, and intermediate results of the pattern matching.
A matching atom is a tuple of a pattern, a matcher, and an expression called \emph{target}.
\texttt{MState} denotes the data constructor for matching states.
\texttt{env} is the environment when the evaluation enters the \texttt{match-all} expression.
A stack of matching atoms contains a single matching atom whose pattern, target and matcher are the arguments of the \texttt{match-all} expression.

\begin{figure}[t]
\centering
\begin{tabular}{m{0.3cm} m{11.8cm}}\
1 & {\footnotesize \begin{verbatim}MState {[<cons $m <cons ,m _>> (multiset integer) {2 8 2}]} env {}\end{verbatim}} \\[-0.7cm] \\\hline \\[-0.7cm]
2 & {\footnotesize \begin{verbatim}
MState {[$m integer 2] [<cons ,m _> (multiset integer) {8 2}]} env {}
MState {[$m integer 8] [<cons ,m _> (multiset integer) {2 2}]} env {}
MState {[$m integer 2] [<cons ,m _> (multiset integer) {2 8}]} env {}
\end{verbatim}} \\[-0.7cm] \\\hline \\[-0.7cm]
3 & {\footnotesize
\begin{verbatim}
MState {[$m something 2] [<cons ,m _> (multiset integer) {8 2}]} env {}
\end{verbatim}} \\[-0.7cm] \\\hline \\[-0.7cm]
4 & {\footnotesize
\begin{verbatim}
MState {[<cons ,m _> (multiset integer) {8 2}]} env {[m 2]}
\end{verbatim}} \\[-0.7cm] \\\hline \\[-0.7cm]
5 & {\footnotesize
\begin{verbatim}
MState {[,m integer 8] [_ (multiset integer) {2}]} env {[m 2]}
MState {[,m integer 2] [_ (multiset integer) {8}]} env {[m 2]}
\end{verbatim}} \\[-0.7cm] \\\hline \\[-0.7cm]
6 & {\footnotesize
\begin{verbatim}
MState {[_ (multiset integer) {8}]} env {[m 2]}
\end{verbatim}} \\[-0.7cm] \\\hline \\[-0.7cm]
7 & {\footnotesize
\begin{verbatim}
MState {[_ something {8}]} env {[m 2]}
\end{verbatim}} \\[-0.7cm] \\\hline \\[-0.7cm]
8 & {\footnotesize
\begin{verbatim}
MState {} env {[m 2]}
\end{verbatim}}
\end{tabular}
  \caption{Reduction path of matching states}
  \label{fig:reductionPath}
\end{figure}

In our proposal, pattern matching is implemented as reductions of matching states.
In a reduction step, the top matching atom in the stack of matching atoms is popped out.
This matching atom is passed to the procedure called \emph{matching function}.
The matching function is a function that takes a matching atom and returns a list of lists of matching atoms.
The behavior of the matching function is controlled by the matcher of the argument matching atom.
We can control the behavior of the matching function by defining matchers properly.
For example, we obtain the following results by passing the matching atom of the initial matching state to the matching function.
\begin{lstlisting}[language=egison]
matchFunction [<cons $m <cons ,m _>> (multiset integer) {2 8 2}] =
  { {[$m integer 2] [<cons ,m _> (multiset integer) {8 2}]}
    {[$m integer 8] [<cons ,m _> (multiset integer) {2 2}]}
    {[$m integer 2] [<cons ,m _> (multiset integer) {2 8}]} }
\end{lstlisting}
\noindent 
Each list of matching atoms is prepended to the stack of the matching atoms.
As a result, the number of matching states increases to three (step 2).
Our pattern-matching system repeats this step until all the matching states vanish.

For simplicity, in the following, we only examine the reduction of the first matching state in step 2.
This matching state is reduced to the matching state shown in step 3.
The matcher in the top matching atom in the stack is changed to \texttt{something} from \texttt{integer}, by definition of \texttt{integer} matcher.
\texttt{something} is the only built-in matcher of our pattern-matching system.
\texttt{something} can handle only wildcards or pattern variables, and is used to bind a value to a pattern variable.
This matching state is then reduced to the matching state shown in step 4.
The top matching atom in the stack is popped out, and a new binding \texttt{[m 2]} is added to the collection of intermediate results.
Only \texttt{something} can append a new binding to the result of pattern matching.

Similarly to the preceding steps, the matching state is then reduced as shown in step 5, and the number of matching states increases to 2.
``\texttt{,m}'' is pattern-matched with \texttt{8} and \texttt{2} by \texttt{integer} matcher in the next step.
When we pattern-match with a value pattern, the intermediate results of the pattern matching is used as an environment to evaluate it.
In this way, ``\texttt{m}'' is evaluated to \texttt{2}.
Therefore, the first matching state fails to pattern-match and vanishes.
The second matching state succeeds in pattern matching and is reduced to the matching state shown in step 6.
In step 7, the matcher is simply converted from \texttt{(multiset integer)} to \texttt{something}, by definition of \texttt{(multiset integer)}.
Finally, the matching state is reduced to the empty collection (step 8).
No new binding is added because the pattern is a wildcard.
When the stack of matching atoms is empty, reduction finishes and the matching patching succeeds for this reduction path.
The matching result \verb|{[m 2]}| is added to the entire result of pattern matching.

We can check the pattern matching for sequential triples and quadruples are also efficiently executed in this algorithm.

\subsection{Pattern Matching with Infinitely Many Results}\label{mechanismInf}

\begin{figure}[t]
  \begin{tabular}{cc}
    \begin{minipage}[t]{0.45\hsize}
      \centering
      \includegraphics[width=4.2cm]{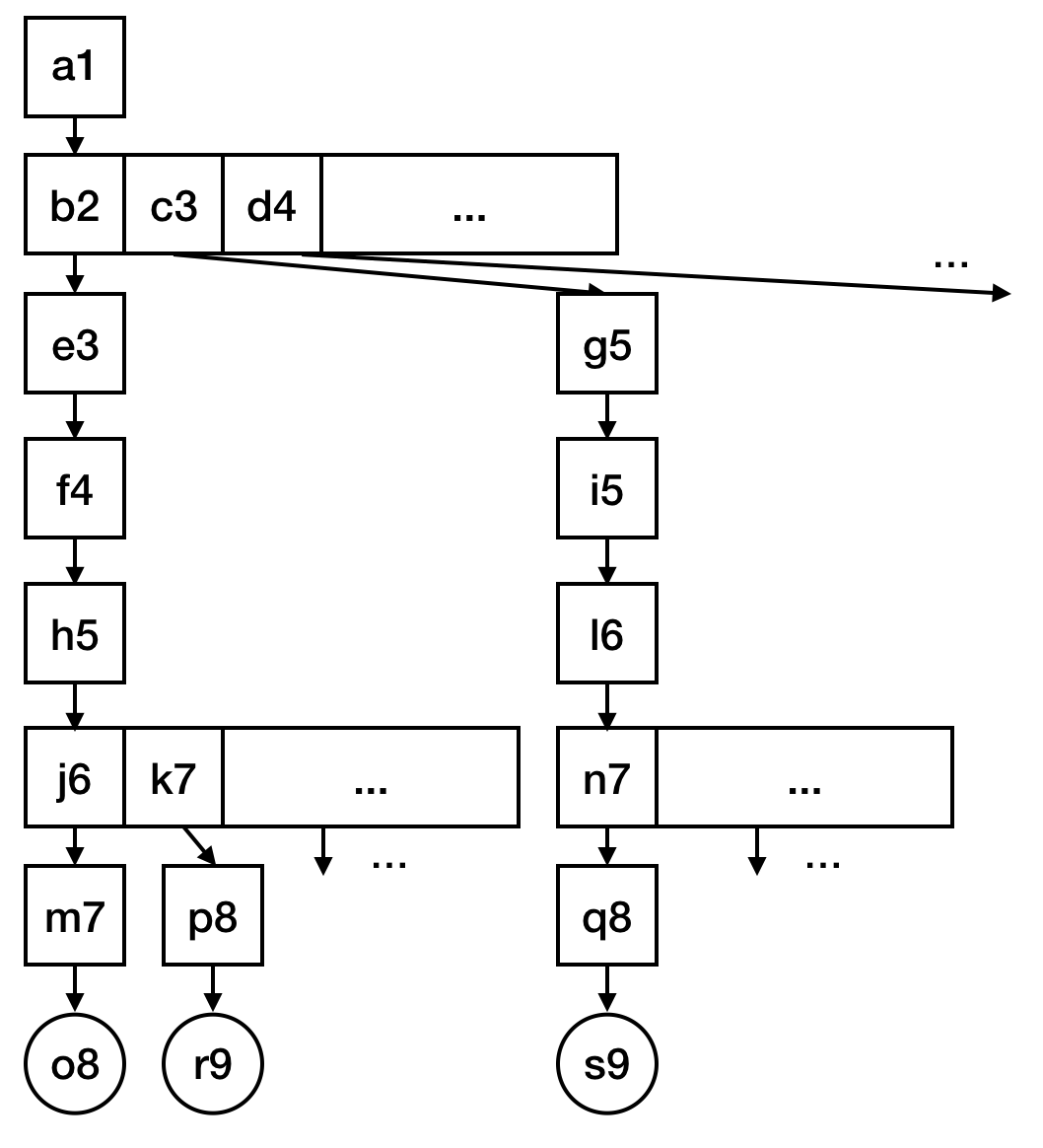}
      \caption{Search tree}
      \label{fig:reduction-tree-1}
    \end{minipage} &
    \begin{minipage}[t]{0.45\hsize}
      \centering
      \includegraphics[width=6.5cm]{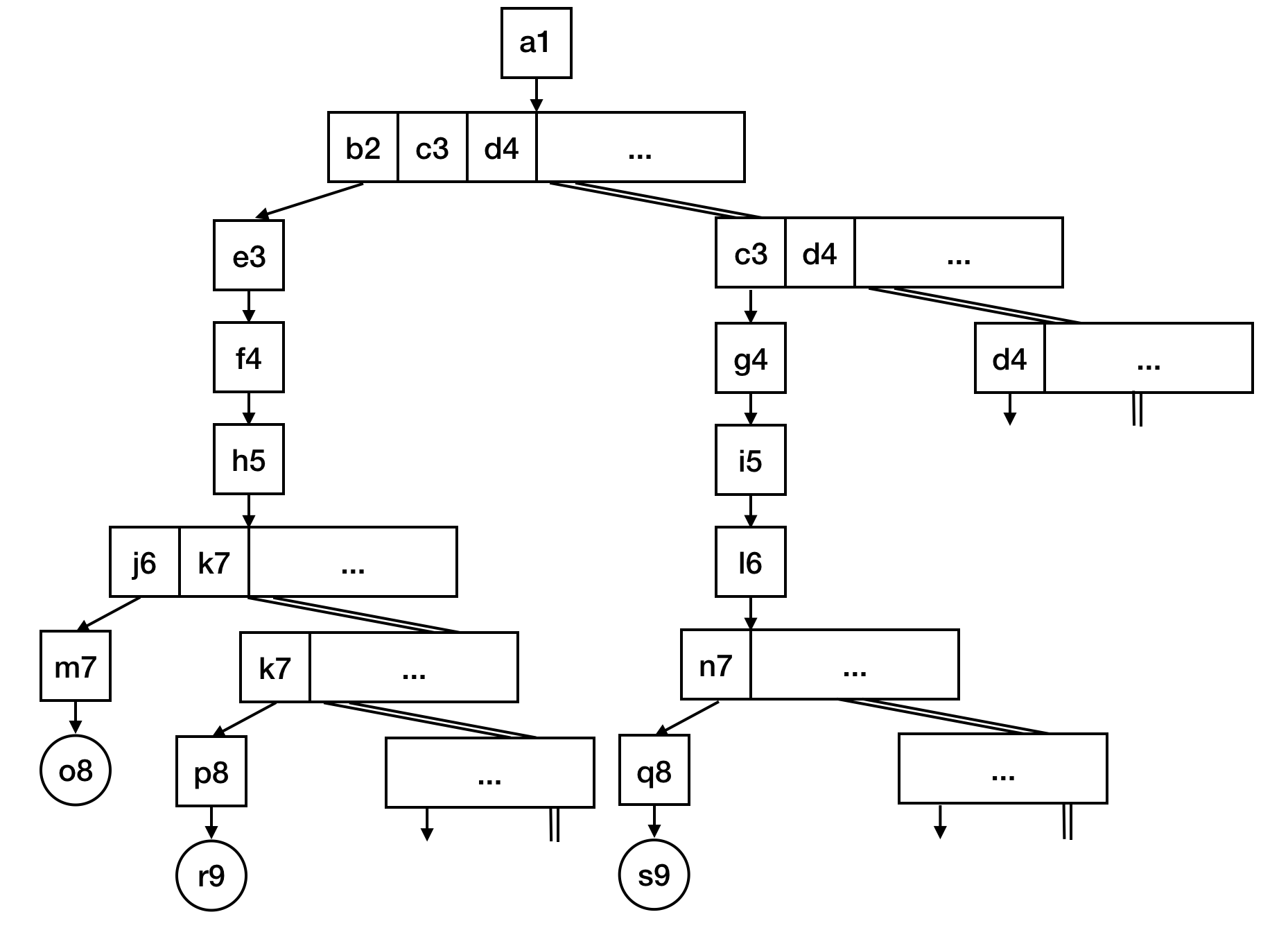}
      \caption{Binary reduction tree}
      \label{fig:reduction-tree-2}
    \end{minipage}
  \end{tabular}
\end{figure}

The proposed pattern-matching system can eventually enumerate all successful matching results when matching results are infinitely many.
It is performed by reducing the matching states in a proper order.
Suppose the following program:
\begin{lstlisting}[language=egison]
(take 8 (match-all nats (set integer) [<cons $m <cons $n _>> [m n]]))
; {[1 1] [1 2] [2 1] [1 3] [2 2] [3 1] [1 4] [2 3]}
\end{lstlisting}
\noindent
Figure~\ref{fig:reduction-tree-1} shows the search tree of matching states when the system executes the above pattern matching expression.
Rectangles represent matching states, and circles represent final matching states of successful pattern matching.
The rectangle at the upper left is the initial matching state.
The rectangles in the second row are the matching states generated from the initial matching state one step.
Circles o8, r9, and s9 correspond to pattern-matching results \verb|{[m 1] [n 1]}|, \verb|{[m 1] [n 2]}|, and \verb|{[m 2] [n 1]}|, respectively.

One issue on naively searching this search tree is that we cannot enumerate all matching states either in depth-first or breadth-first manners.
The reason is that widths and depths of the search tree can be infinite.
Widths can be infinite because a matching state may generate infinitely many matching states (e.g., the width of the second row is infinite), and depths can be infinite when we extend the language with a notion such as recursively defined patterns~\cite{egi2014non}.

To resolve this issue, we reshape the search tree into a \emph{reduction tree} as presented in Figure~\ref{fig:reduction-tree-2}.
A node of a reduction tree is a list of matching states, and a node has at most two child nodes, left of which is the matching states generated from the head matching state of the parent, and right of which is a copy of the tail part of the parent matching states.
At each reduction step, the system has a list of nodes.
Each row in Figure~\ref{fig:reduction-tree-2} denotes such a list.
One reduction step in our system proceeds in the following two steps.
First, for each node, it generates a node from the head matching state.
Then, it constructs the nodes for the next step by collecting the generated nodes and the copies of the tail parts of the nodes.
The index of each node denotes the depth in the tree the node is checked at.
Since widths of the tree are at most $2^n$ for some $n$ at any depth, all nodes can be assigned some finite number, which means all nodes in the tree are eventually checked after a finite number of reduction steps.

We adopt breadth-first search strategy as the default traverse method because there are cases that breadth-first traverse can successfully enumerate all pattern-matching results while depth-first traverse fails to do so when we handle pattern matching with infinitely many results.
However, of course, when the size of the reduction tree is finite, the space complexity for depth-first traverse is less expensive.
Furthermore, there are cases that the time complexity for depth-first traverse is also less expensive when we extract only the first several successful matches.
Therefore, to extend the range of algorithms we can express concisely with pattern matching keeping efficiency, providing users with a method for switching search strategy of reduction trees is important.
We leave further investigation of this direction as as interesting future work.

\section{User Defined Matchers}\label{matcher}

This section explains how to define matchers.

\subsection{Matcher for Unordered Pairs}\label{unordered-pair}

We explain how the \texttt{unordered-pair} matcher shown in Section~\ref{egison-extensible} works.
\texttt{unordered-pair} is defined as a function that takes and returns a matcher to specify how to pattern-match against the elements of a pair.
\texttt{matcher} takes matcher clauses.
A matcher clause is a triple of a primitive-pattern pattern, next-matcher expressions, and primitive-data-match clauses.
The formal syntax of the \texttt{matcher} expression is found in Figure~\ref{fig:syntax} in Section~\ref{formal-semantics}.

\texttt{unordered-pair} has two matcher clauses.
The primitive-pattern pattern of the first matcher clause is \verb|<pair $ $>|.
This matcher clause defines the interpretation of \texttt{pair} pattern.
\texttt{pair} takes two pattern holes \verb|$|.
It means that it interprets the first and second arguments of \texttt{pair} pattern by the matchers specified by the next-matcher expression.
In this example, since the next-matcher expression is \verb|[a a]|, both of the arguments of \texttt{pair} are pattern-matched using the matcher given by \verb|a|.
The primitive-data-match clause of the first matcher clause is \verb|{[<Pair $x $y> {[x y] [y x]}]}|.
\texttt{<Pair \$x \$y>} is pattern-matched with the target datum such as \texttt{<Pair 2 5>},
and \texttt{\$x} and \texttt{\$y} is matched with \texttt{2} and \texttt{5}, respectively.
The primitive-data-match clause returns \verb|{[2 5] [5 2]}|.
A primitive-data-match clause returns a collection of \textit{next-targets}.
This means the patterns ``\texttt{,5}'' and \texttt{\$x} are matched with the targets \texttt{2} and \texttt{5}, or \texttt{5} and \texttt{2} using the \texttt{integer} matcher in the next step, respectively.
Pattern matching of primitive-data-patterns is similar to pattern matching against algebraic data types in ordinary functional programming languages.
As a result, the first matcher clause works in the matching function as follows.
\begin{lstlisting}[language=egison]
matchFunction [<pair $x $y> (unordered-pair integer) <Pair 2 5>] =
  { {[$x integer 2] [$y integer 5]} {[$x integer 5] [$y integer 2]} }
\end{lstlisting}

The second matcher clause is rather simple; this matcher clause simply converts the matcher of the matching atom to the \verb|something| matcher.

\subsection{Case Study: Matcher for Multisets}\label{multiset-matcher}

As an example of how we can implement matchers for user-defined non-free data types, we show the definition of \texttt{multiset} matcher.
We can define it simply by using the \texttt{list} matcher.
\texttt{multiset} is defined as a function that takes and returns a matcher.
\begin{lstlisting}[language=egison]
(define $multiset
  (lambda [$a]
    (matcher
      {[<nil> [] {[{} {[]}] [_ {}]}]
       [<cons $ $> [a (multiset a)]
        {[$tgt (match-all tgt (list a)
                 [<join $hs <cons $x $ts>>
                  [x (append hs ts)]])]}]
       [,$val []
        {[$tgt (match [val tgt] [(list a) (multiset a)]
                 {[[<nil> <nil>] {[]}]
                  [[<cons $x $xs> <cons ,x ,xs>] {[]}]
                  [[_ _] {}]})]}]
       [$ [something] {[$tgt {tgt}]}]})))
\end{lstlisting}
The \texttt{multiset} matcher has four matcher clauses.
The first matcher clause handles the \texttt{nil} pattern, and it checks if the target is an empty collection.
The second matcher clause handles the \texttt{cons} pattern.
The \texttt{match-all} expression is effectively used to destruct a collection in the primitive-data-match clause.
Because the \texttt{join} pattern in the \texttt{list} matcher enumerates all possible splitting pairs of the given list, \texttt{match-all} lists up all possible consing pairs of the target expression.
The third matcher clause handles value patterns.
``\verb|,$val|'' is a value-pattern pattern that matches with a value pattern.
This matcher clause checks if the content of a value pattern (bound to \texttt{val}) is equal to the target (bound to \texttt{tgt}) as multisets.
Note that the definition involves recursions on the \texttt{multiset} matcher itself.
The fourth matcher clause is completely identical to \texttt{unordered-pair} and \texttt{integer}.


\subsection{Value-pattern Patterns and Predicate Patterns}\label{integer}

We explain the generality of our extensible pattern-matching framework taking examples from the \texttt{integer} matcher.
How to implement value patterns and predicate patterns in our language is shown.
\begin{lstlisting}[language=egison]
(define $integer
  (matcher {[,$n [] {[$tgt (if (eq? tgt n) {[]} {})]}]
            [<lt ,$n> [] {[$tgt (if (lt? tgt n) {[]} {})]}]
            [$ [something] {[$tgt {tgt}]}]}))
\end{lstlisting}
Value patterns are patterns that successfully match if the target expression is equal to some fixed value.
For example, \verb|,5| only matches with \verb|5| if we use \texttt{integer} matcher.
The first matcher clause in the above definition exists to implement this.
The primitive-pattern pattern of this clause is \verb|,$n|, which is a value-pattern pattern that matches with value patterns.
The next-matcher expression is an empty tuple because no pattern hole \verb|$| is contained.
If the target expression \texttt{tgt} and the content of the value pattern \texttt{n} are equal, the primitive-data-match clause returns a collection consisting of an empty tuple, which denotes success.
Otherwise, it returns an empty collection, which denotes failure.

Predicate patterns are patterns that succeed if the target expression satisfies some fixed predicate.
Predicate patterns are usually implemented as a built-in feature, such as pattern guards, in ordinary programming languages.
Interestingly, we can implement this on top of our pattern-matching framework.
The second matcher clause defines a predicate pattern which succeeds if the target integer is less than the content of the value pattern \verb|n|.
A technique similar to the first clause is used.

\section{Formal Semantics}\label{formal-semantics}

\begin{figure}[!t]
\scriptsize
\begin{multicols}{2}
\noindent
\begin{align*}
M &::= x \mid c \mid \texttt{(lambda [\$$x$ $\cdots$] $M$)} \mid \texttt{($M$ $M$ $\cdots$)} \\
&\mid [M \cdots] \mid \{M \cdots\} \mid \texttt{<$C$ $M$ $\cdots$>} \\
&\mid \texttt{(match-all $M$ $M$ [$p$ $M$])} \\
&\mid \texttt{(match $M$ $M$ \{[$p$ $M$] $\cdots$\})} \\
&\mid \texttt{something} \mid \texttt{(matcher \{$\phi$ $\cdots$\})}
\end{align*}
\columnbreak
\begin{align*}
p &::= \texttt{_} \mid \texttt{\$$x$} \mid \texttt{,$M$} \mid \texttt{<$C$ $p$ $\cdots$>} \\
\phi &::= \texttt{[$pp$ $M$ \{[$dp$ $M$] $\cdots$\}]} \\
pp &::= \texttt{\$} \mid \texttt{,\$$x$} \mid \texttt{<$C$ $pp$ $\cdots$>} \\
dp &::= \texttt{\$$x$} \mid \texttt{<$C$ $dp$ $\cdots$>}
\end{align*}
\end{multicols}
  \caption{Syntax of our language}
  \label{fig:syntax}
\end{figure}

In this section, we present the syntax and big-step semantics of our language (Fig.~\ref{fig:syntax} and~\ref{fig:formal-semantics}).
We use metavariables $x,y,z,\ldots$, $M,N,L,\ldots$, $v,\ldots$, and $p,\ldots$ for variables, expressions, values, and patterns respectively.
In Fig.~\ref{fig:syntax}, $c$ denotes a constant expression and $C$ denotes a data constructor name.
$X \cdots$ in Fig.~\ref{fig:syntax} means a finite list of $X$.
The syntax of our language is similar to that of the Lisp language.
As explained in Section~\ref{egison-expr}, \texttt{[$M$ $\cdots$]}, \texttt{\{$M$ $\cdots$\}}, and \texttt{<$C$ $M$ $\cdots$>} denote tuples, collections, and data constructions.
All formal arguments are decorated with the dollar mark.
$\phi$, $pp$ and $dp$ are called matcher clauses, primitive-pattern patterns and primitive-data patterns respectively.

\begin{figure}[!t]
  \scriptsize
  \begin{gather*}
  \intertext{Evaluation of \texttt{matcher} and \texttt{match-all}:}
    \AXC{}
    \UIC{$\ev{\Gamma}{\texttt{(matcher $\texttt{[$pp_i$ $M_i$ $\texttt{[$dp_{ij}$ $N_{ij}$]}_j$]}_i$)}} ([pp_i, M_i, [dp_{ij}, N_{ij}]_j]_i,\Gamma)$}
    \DP
    \\
    \AXC{$\ev{\Gamma}{M} v$}
    \AXC{$\ev{\Gamma}{N} m$}
    \AXC{$[[[\matom{p}{m}{v}],\Gamma,\emptyset]] \Rrightarrow [\Delta_i]_i $}
    \AXC{$\ev{\Gamma \cup \Delta_i}{L} v_i \quad (\forall i)$}
    \QIC{$\ev{\Gamma}{\texttt{(match-all $M$ $N$ [$p$ $L$])}} [v_i]_i$}
    \DP
  \intertext{Matching states:}
    \AXC{}
    \UIC{$\epsilon \rightarrow \none, \none, \none$}
    \DP
    \quad
    \AXC{}
    \UIC{$(\epsilon, \Gamma, \Delta):\vec{s} \rightarrow (\some \Delta), \none, (\some \vec{s})$}
    \DP
    \\
    \AXC{$\mfun{p}{\Gamma \cup \Delta}{m}{v} [\vec{a}_i]_i, \Delta'$}
    \UIC{$((\matom{p}{m}{v}) \cons \vec{a}, \Gamma, \Delta):\vec{s} \rightarrow \none, (\some [\vec{a}_i + \vec{a}, \Gamma, \Delta \cup \Delta']_i), (\some \vec{s})$}
    \DP
    \\
    \AXC{$\vec{s}_i \rightarrow \opt \Gamma_i, \opt \vec{s'}_i, \opt \vec{s''}_i \quad (\forall i)$}
    \UIC{$[\vec{s}_i]_i \Rightarrow \sum_i (\opt \Gamma_i), \sum_i (\opt \vec{s'}_i) + \sum_i (\opt \vec{s''}_i)$}
    \DP
    \qquad
    \AXC{$\mathstrut$}
    \UIC{$\epsilon \Rrightarrow \epsilon$}
    \DP
    \quad
    \AXC{$\vec{\vec{s}} \Rightarrow \vec{\Gamma}, \vec{\vec{s'}}$}
    \AXC{$\vec{\vec{s'}} \Rrightarrow \vec{\Delta}$}
    \BIC{$\vec{\vec{s}} \Rrightarrow \vec{\Gamma} + \vec{\Delta}$}
    \DP
  \intertext{Matching atoms:}
    \AXC{$\mathstrut$}
    \UIC{$\mfun{\texttt{\$x}}{\Gamma}{\texttt{something}}{v} [\epsilon], \set{x \mapsto v}$}
    \DP
    \quad
    \AXC{$\ppm{pp}{\Gamma}{p} \textbf{fail}$}
    \AXC{$\mfun{p}{\Gamma}{(\vec{\phi},\Delta)}{v} \vec{\vec{a}}, \Gamma'$}
    \BIC{$\mfun{p}{\Gamma}{((pp,M,\vec{\sigma}) \cons \vec{\phi},\Delta)}{v} \vec{\vec{a}}, \Gamma'$}
    \DP
    \\
    \AXC{$\ppm{pp}{\Gamma}{p} [p'_i]_i, \Delta'$}
    \AXC{$\pdm{dp}{v} \textbf{fail}$}
    \AXC{$\mfun{p}{\Gamma}{((pp,M,\vec{\sigma}) \cons \vec{\phi},\Delta)}{v} \vec{\vec{a}}, \Gamma'$}
    \TIC{$\mfun{p}{\Gamma}{((pp,M,(dp,N) \cons \vec{\sigma}) \cons \vec{\phi},\Delta)}{v} \vec{\vec{a}}, \Gamma'$}
    \DP
    \\
    \AXC{$\ppm{pp}{\Gamma}{p} [p'_j]_j, \Delta'$}
    \AXC{$\pdm{dp}{v} \Delta''$}
    \AXC{$\ev{\Delta \cup \Delta' \cup \Delta''}{N} [[v'_{ij}]_j]_i$}
    \AXC{$\ev{\Delta}{M} [m'_j]_j$}
    \QIC{$\mfun{p}{\Gamma}{((pp,M,(dp,N) \cons \vec{\sigma}) \cons \vec{\phi},\Delta)}{v} [[\matom{p'_j}{m'_j}{v'_{ij}}]_j]_i, \emptyset$}
    \DP
  \intertext{Pattern matching on patterns:}
    \AXC{$\mathstrut$}
    \UIC{$\ppm{\texttt{\$}}{\Gamma}{p} [p], \emptyset$}
    \DP
    \quad
    \AXC{$\ev{\Gamma}{M} v$}
    \UIC{$\ppm{\texttt{,\$y}}{\Gamma}{\texttt{,$M$}} \epsilon, \set{y \mapsto v}$}
    \DP
    \quad
    \AXC{$\ppm{pp_i}{\Gamma}{p_i} \vec{p}_i, \Gamma_i \quad (\forall i)$}
    \UIC{$\ppm{\texttt{<C $pp_1 \ldots pp_n$>}}{\Gamma}{\texttt{<C $p_1 \ldots p_n$>}} \sum_i \vec{p}_i, \bigcup_i \Gamma_i$}
    \DP
  \intertext{Pattern matching on data:}
    \AXC{$\mathstrut$}
    \UIC{$\pdm{\texttt{\$z}}{v} \set{z \mapsto v}$}
    \DP
    \qquad
    \AXC{$\pdm{dp_i}{v_i} \Gamma_i \quad (\forall i)$}
    \UIC{$\pdm{\texttt{<C $dp_1 \ldots dp_n$>}}{\texttt{<C $v_1 \ldots v_n$>}} \bigcup_i \Gamma_i$}
    \DP
  \end{gather*}
  \caption{Formal semantics of our language}
  \label{fig:formal-semantics}
\end{figure}

In Fig.~\ref{fig:formal-semantics}, the following notations are used.
We write $[a_i]_i$ to mean a list $[a_1, a_2, \ldots]$.
Similarly, $[[a_{ij}]_j]_i$ denotes $[[a_{11},a_{12},\ldots],[a_{21},a_{22},\ldots],\ldots]$, but each list in the list may have different length.
List of tuples $[(a_1,b_1),(a_2,b_2),\ldots]$ may be often written as $[a_i,b_i]_i$ instead of $[(a_i,b_i)]_i$ for short.
Concatenation of lists $l_1,l_2$ are denoted by $l_1 + l_2$, and $a \cons l$ denotes $[a] + l$ (adding at the front).
$\epsilon$ denotes the empty list.
In general, $\vec{x}$ for some metavariable $x$ is a metavariable denoting a list of what $x$ denotes.
However, we do \emph{not} mean by $\vec{x}_i$ the $i$-th element of $\vec{x}$; if we write $[\vec{x}_i]_i$, we mean a list of a list of $x$.
$\Gamma, \Delta, \ldots$ denote variable assignments, i.e., partial functions from variables to values.

Our language has some special primitive types: matching atoms $a, \ldots$, matching states $s,\ldots$, primitive-data-match clauses $\sigma, \ldots$, and matchers $m,\ldots$.
A matching atom consists of a pattern $p$, a matcher $m$, and a value $v$, and written as $\matom{p}{m}{v}$.
A matching state is a tuple of a list of matching atoms and two variable assignments.
A primitive-data-match clause is a tuple of a primitive-data pattern and an expression, and a matcher clause is a tuple of a primitive-pattern pattern, an expression, and a list of data-pattern clauses.
A matcher is a pair containing a list of matcher clauses and a variable assignment.
Note that matchers, matching states, etc.\, are all values.

Evaluation results of expressions are specified by the judgment $\ev{\Gamma}{e} \vec{v}$, which denotes given a variable assignment $\Gamma$ and an expression $e$ one gets a list of values $\vec{v}$.
In the figure, we only show the definition of evaluation of \texttt{matcher} and \texttt{match-all} expressions (other cases are inductively defined as usual).
The definition of \texttt{match-all} relies on another type of judgment $\vec{\vec{s}} \Rrightarrow \vec{\Gamma}$, which defines how the search space is examined.
$\Rrightarrow$ is inductively defined using $\vec{\vec{s}} \Rightarrow \vec{\Gamma}, \vec{\vec{s'}}$, which is again defined using $\vec{s} \rightarrow \opt \Gamma, \opt \vec{s'}, \opt \vec{s''}$.
In their definitions, we introduced notations for (meta-level) option types.
$\none$ and $\some{x}$ are the constructors of the option type, and $\opt x$ is a metavariable for an optional value (possibly) containing what the metavariable $x$ denotes.
$\sum_i (\opt x_i)$ creates a list by collecting all the valid (non-$\none$) $x_i$ preserving the order.

$\mfun{p}{\Gamma}{m}{v} \vec{\vec{a}},\Delta$ is a 6-ary relation.
One reads it ``performing pattern matching on $v$ against $p$ using the matcher $m$ under the variable assignment $\Gamma$ yields the result $\Delta$ and continuation $\vec{\vec{a}}$.''
The result is a variable assignment because it is a result of unifications.
$\vec{\vec{a}}$ being empty means the pattern matching failed.
If $[\epsilon]$ is returned as $\vec{\vec{a}}$, it means the pattern matching succeeded and no further search is necessary.
As explained in Section~\ref{matcher}, one needs to pattern-match patterns and data to define user-defined matchers.
Their formal definitions are given by judgments $\ppm{pp}{\Gamma}{p} \vec{p'}, \Delta$ and $\pdm{dp}{v} \Gamma$.

\section{Conclusion}\label{conclusion}

We designed a user-customizable efficient non-linear pattern-matching system by regarding pattern matching as reduction of matching states that have a stack of matching atoms and intermediate results of pattern matching.
This system enables us to concisely describe a wide range of programs, especially when non-free data types are involved.
For example, our pattern matching architecture is useful to implement a computer algebra system because it enables us to directly pattern-match mathematical expressions and rewrite them.

The major significance of our pattern matching system is that it greatly improves the expressivity of the programming language by allowing programmers to freely extend the process of pattern matching by themselves.
Furthermore, in the general cases, use of the \texttt{match} expression will be as readable as that in other general-purpose programming languages.
Although we consider that the current syntax of matcher definition is already clean enough, we leave further refinement of the syntax of our surface language as future work.

We believe the direct and concise representation of algorithms enables us to implement really new things that go beyond what was considered practical before.
We hope our work will lead to breakthroughs in various fields.

\section*{Acknowledgments}
We thank Ryo Tanaka, Takahisa Watanabe, Kentaro Honda, Takuya Kuwahara, Mayuko Kori, and Akira Kawata for their important contributions to implement the interpreter.
We thank Michal J. Gajda, Yi Dai, Hiromi Hirano, Kimio Kuramitsu, and Pierre Imai for their helpful feedback on the earlier versions of the paper.
We thank Masami Hagiya, Yoshihiko Kakutani, Yoichi Hirai, Ibuki Kawamata, Takahiro Kubota, Takasuke Nakamura, Yasunori Harada, Ikuo Takeuchi, Yukihiro Matsumoto, Hidehiko Masuhara, and Yasuhiro Yamada for constructive discussion and their continuing encouragement.

\bibliographystyle{splncs04}
\bibliography{egison}


\end{document}